\documentclass[twocolumn,showpacs,preprintnumbers,showkeys,superscriptaddress]{revtex4}
\usepackage{amssymb}
\usepackage{graphicx}
\usepackage{dcolumn}
\usepackage{bm}
\usepackage{appendix}

\def\eqref#1{Eq.~(\ref{eq:#1})}

\begin{document}

\title{Practical Calculation Scheme for Generalized Seniority}
\author{L. Y. Jia}  \email{liyuan.jia@usst.edu.cn}
\affiliation{Department of Physics, University of Shanghai for
Science and Technology, Shanghai 200093, P. R. China}
\affiliation{Department of Physics, Hebei Normal University,
Shijiazhuang, Hebei 050024, P. R. China}

\date{\today}

\begin{abstract}

We propose a scheme or procedure for doing practical calculations
with generalized seniority. It reduces the total computing time by
calculating and storing in advance a set of intermediate quantities,
taking advantage of the memory capability of modern computers. The
requirements and performance of the algorithm are analyzed in
detail.

\end{abstract}


\vspace{0.4in}

\maketitle

\section{Introduction}

Generalized seniority has long been introduced \cite{Talmi_1971,
Shlomo_1972, Allaart_1988, Gambhir_1969} in nuclear physics as an
effective truncation scheme for the nuclear shell model. In the
presence of strong pairing correlations, the nucleons form pairs and
the ground state of an even-even nucleus is usually well
approximated by a pair condensate. For the low-lying states the
number of broken pairs should be small because naively breaking each
pair costs about $2$ MeV in energy (pairing energy). Consequently,
generalized seniority $S$ is introduced as the number of particles
not participating in the coherent pair condensate (unpaired
particles), and it is usually a good approximation to truncate the
full many-body space to the one consisting of the states with low
seniority.

In the literature there are many ways to calculate the matrix
elements of operators between states with fixed seniority. Explicit
expressions in various forms have been derived for cases of the
lowest seniorities \cite{Gambhir_1969, Gambhir_1971,
Bonsignori_1978, Pittel_1982, Scholten_1983, Frank_1982,
Isacker_1986, IBM_book, Mizusaki_1996} and applied to realistic
nuclei \cite{Scholten_1983_2, Bonsignori_1985, Engel_1989,
Monnoye_2002}, but for higher seniority these expressions rapidly
become cumbersome and have only formal meanings. Recently recursive
relations for the matrix elements were derived \cite{Luo_2011} using
the angular-momentum coupled version of the Wick's theorem
\cite{Chen_1993_NPA1,Chen_1993_NPA2}, however in realistic
calculations these relations may become very time consuming and up
to now the method has only been carried out for $S \le 2$ (one
broken pair for each species of nucleons)
\cite{Caprio_2012_PRC,Caprio_2012_JPG}.

Computers have enjoyed rapid growth recently, in both computing
speed and data storage capacity. Nowadays it is common to have
several gigabyte memory in one's laptop, and several terabyte memory
at a workstation. The aim of this work is to propose an algorithm or
procedure of generalized seniority that reduces the total time costs
by utilizing the huge memories. The matrix elements of operators
between seniority states are calculated in two steps. We first
compute and store in memory the ``density matrix'' on the pair
condensate that characterizes the properties of the latter. Then the
matrix elements of operators are expressed in terms of the ``density
matrix'' through simple relations. In Sec. \ref{Sec_pair} we
introduce the ``density matrix'' and derive recursive relations for
its calculation. The simple routine is given to express the matrix
elements of operators between seniority states in terms of the
``density matrix''. Section \ref{Sec_shell} is devoted to the
seniority truncation of the shell model. We formulate the procedure
explicitly, the requirements (speed and memory) and performance of
the algorithm are analyzed in detail. Finally in Sec. \ref{Sec_sum}
we summarize the work.

\section{Matrix Elements on the Pair Condensate  \label{Sec_pair}}

In this section we consider the matrix elements of operators on the
pair condensate. The pair-creation operator
\begin{eqnarray}
P_1^\dagger = a_1^\dagger a_{\tilde{1}}^\dagger  \label{P1_dag}
\end{eqnarray}
creates a pair of particles on the single-particle level $|1\rangle$
and its time-reversed partner $|\tilde{1}\rangle$
($|\tilde{\tilde{1}}\rangle = - |1\rangle$). The coherent
pair-creation operator
\begin{eqnarray}
P^\dagger = \sum_\alpha v_\alpha P_\alpha^\dagger \label{P_dag}
\end{eqnarray}
creates a pair of particles coherently distributed with structure
coefficients $v_\alpha$ over the entire single-particle space. In
Eq. (\ref{P_dag}) the summation index $\alpha$ is the ``pair index''
that runs over only half of the single-particle space ($P_{1} =
P_{\tilde{1}}$). In the presence of pairing correlations, the
seniority zero state of the $2N$-particle system is
\begin{eqnarray}
|\phi_N\rangle = \frac{1}{\sqrt{\chi_{N}}} (P^\dagger)^{N} |0\rangle
, \label{gs}
\end{eqnarray}
where
\begin{eqnarray}
\chi_{N} = \langle 0 | P^N (P^\dagger)^{N} | 0 \rangle \label{chi_N}
\end{eqnarray}
is the normalization factor. In addition, we introduce the
pair-transfer amplitudes
\begin{eqnarray}
t_{\alpha_1 \alpha_2 ... \alpha_p;\beta_1 \beta_2 ... \beta_q}^{M} =
\nonumber \\
\langle 0 | P^{M-p} P_{\alpha_1} P_{\alpha_2} ... P_{\alpha_p}
P_{\beta_1}^\dagger P_{\beta_2}^\dagger ... P_{\beta_q}^\dagger
(P^\dagger)^{M-q} | 0 \rangle ,  \label{t_ij}
\end{eqnarray}
where \emph{by definition all the ``pair indices'' $\alpha_1, ...,
\alpha_p, \beta_1, ..., \beta_q$ are different}. The real number
$t_{\alpha_1 \alpha_2 ... \alpha_p;\beta_1 \beta_2 ... \beta_q}^{M}$
is symmetric under permutations within $\alpha$ indices or $\beta$
indices; and $t_{\alpha_1 \alpha_2 ... \alpha_p;\beta_1 \beta_2 ...
\beta_q}^{M} = t_{\beta_1 \beta_2 ... \beta_q;\alpha_1 \alpha_2 ...
\alpha_p}^{M}$. The normalization $\chi_N$ defined in Eq.
(\ref{chi_N}) is the special case of Eq. (\ref{t_ij}) when all the
$\alpha$ and $\beta$ indices are missing: $\chi_N = t_{;}^{N}$.

Recursive relations for $t$ (\ref{t_ij}) exist because operators
$P_1^\dagger$ (\ref{P1_dag}), $P_1 = (P_1^\dagger)^\dagger =
a_{\tilde{1}} a_1$, and $\hat{N}_1 = \frac{1}{2} ( a_1^\dagger a_1 +
a_{\tilde{1}}^\dagger a_{\tilde{1}} )$ form a closed algebra:
\begin{eqnarray}
[P_1 , P_1^\dagger] = 1 - 2 \hat{N}_1 ,~~ [\hat{N}_1, P_1^\dagger] =
P_1^\dagger .  \label{algebra}
\end{eqnarray}
From Eqs. (\ref{P_dag}) and (\ref{algebra}) it is easy to derive the
identity
\begin{eqnarray}
P_\alpha (P^\dagger)^N | 0 \rangle =  \nonumber \\
v_\alpha N (P^\dagger)^{N-1} | 0 \rangle - (v_\alpha)^2 N (N-1)
P_\alpha^\dagger (P^\dagger)^{N-2} | 0 \rangle , \nonumber
\end{eqnarray}
and consequently the recursive relations for the quantity $t$
(\ref{t_ij}),
\begin{eqnarray}
t_{\alpha_1 \alpha_2 ... \alpha_p;\beta_1 ... \beta_q}^{M} =
v_{\alpha_p} (M-q) t_{\alpha_1 \alpha_2 ...
\alpha_{p-1};\beta_{1} ... \beta_{q}}^{M-1}  \nonumber \\
- (v_{\alpha_p})^2 (M-q) (M-q-1) t_{\alpha_1 \alpha_2 ...
\alpha_{p-1}; \alpha_p \beta_1 ... \beta_q}^{M-1} .  \label{t_rec}
\end{eqnarray}
The simplest case of Eq. (\ref{t_rec}) gives the recursive relation
for the one-pair transfer amplitudes when there is only one $\alpha$
subscript on $t$,
\begin{eqnarray}
t_{\alpha;}^{M} = \langle 0 | P^{M-1} P_{\alpha} (P^\dagger)^{M} | 0
\rangle =  \nonumber \\
v_{\alpha} M \chi_{M-1} - (v_{\alpha})^2 M
(M-1) t_{\alpha;}^{M-1} ,  \label{t1_rec}
\end{eqnarray}
which is Eq. (22) in Ref. \cite{Jia_4}. The normalization
(\ref{chi_N}) is calculated as
\begin{eqnarray}
\chi_{N} = \sum_\alpha v_\alpha t^N_{\alpha;} .  \label{chi_rec}
\end{eqnarray}

The most general operator is written schematically as a product of
single-particle annihilation and creation operators, its matrix
element on the pair condensate (\ref{gs}) is
\begin{eqnarray}
\langle 0 | P^{M} a_{i_1} a_{i_2} ... a_{i_p} a_{j_1}^\dagger
a_{j_2}^\dagger ... a_{j_q}^\dagger (P^\dagger)^{N} | 0 \rangle ,
\label{me_general}
\end{eqnarray}
where $i_1, ..., i_p, j_1, ..., j_q$ are single-particle indices
that take values from the entire single-particle space (both
$|1\rangle$ and $|\tilde{1}\rangle$ are allowed), and $2 M + p = 2 N
+ q$ guarantees particle-number conservation. In order for the
matrix element (\ref{me_general}) to be nonzero, the indices $i_1,
i_2, ..., i_p$ and $j_1, j_2, ..., j_q$ must differ in time-reversed
pairs, because in $(P^\dagger)^N | 0 \rangle$ and $\langle 0 | P^M$
the single-particle levels are occupied in time-reversed pairs. An
example is
\begin{widetext}
\begin{eqnarray}
\langle 0 | P^{N-1} a_{\tilde{1}} a_{1} a_{\tilde{2}} a_{2}
a_{\tilde{3}} a_{3} a_{\tilde{4}} a_{4} a_{5} a_{6} a_{7}
a_7^\dagger a_6^\dagger a_5^\dagger a_{3}^\dagger
a_{\tilde{3}}^\dagger a_{1}^\dagger a_{\tilde{1}}^\dagger
a_{8}^\dagger a_{\tilde{8}}^\dagger (P^\dagger)^{N} | 0 \rangle
\nonumber \\
= \langle 0 | P_1 P_1^\dagger | 0 \rangle \langle 0 | P_3
P_3^\dagger | 0 \rangle \langle 0 | a_{5} a_{6} a_{7} a_7^\dagger
a_6^\dagger a_5^\dagger | 0 \rangle \langle 0^{[1,3,5,6,7]} |
P^{N-1} P_2 P_4 P_{8}^\dagger (P^\dagger)^{N} | 0^{[1,3,5,6,7]}
\rangle = t^{N+1 {[1,3,5,6,7]}}_{2,4;8} ,  \label{me_example}
\end{eqnarray}
\end{widetext}
where $| 0^{[1,3,5,6,7]} \rangle$ represents a subspace of the
original single-particle space, by removing the single-particle
levels $1, \tilde{1}, 3, \tilde{3}, 5, \tilde{5}, 6, \tilde{6}, 7,
\tilde{7}$ from the latter. This is the Pauli blocking effect;
because operators $P_1^\dagger$, $P_3^\dagger$, and $a_7^\dagger
a_6^\dagger a_5^\dagger$ ($P_1$, $P_3$, and $a_5 a_6 a_7$) could be
moved to the rightmost (leftmost) side and their effects were simply
blocking the corresponding pairs of single-particle levels.
Similarly, $t^{N+1 {[1,3,5,6,7]}}_{2,4;8}$ is defined as the
pair-transfer amplitude in this restricted subspace. A formal
analytical expression could be written down involving complicated
Kronecker delta functions, but here we are content with the
programmable routine described in Eq. (\ref{me_example}).

The matrix element of an arbitrary operator $O$ between states with
fixed seniority could be written in the form (\ref{me_general}): the
product $a_{i_1} ... a_{i_p} a_{j_1}^\dagger ... a_{j_q}^\dagger$
consists of the operator $O$ and the unpaired particles. From Eq.
(\ref{me_example}) we see that the matrix element of the form
(\ref{me_general}) boils down to the pair-transfer amplitudes $t$
introduced in Eq. (\ref{t_ij}), \emph{calculated in the original
single-particle space and its subspaces}. These $t$'s play the role
of ``density matrix'' for the pair condensate. They are the
intermediate quantities that appear repeatedly in the calculation
and we would like to compute and store in advance to reduce the
total time costs. In the next section we consider whether this is
possible for realistic calculations within modern computers.

\section{Seniority Truncation of Shell Model  \label{Sec_shell}}

For simplicity we consider semi-magic even-even nuclei that have
only one kind of active nucleons. The matrix elements of a two-body
Hamiltonian between states with fixed seniority $2(\nu - \mu)$ and
$2\nu$ are schematically written as
\begin{eqnarray}
\langle 0 | P^{N+\mu} \underbrace{a a ... a}_{2(\nu - \mu)}
\underbrace{(a a a^\dagger a^\dagger)}_{H} \underbrace{a^\dagger
a^\dagger ... a^\dagger}_{2\nu} (P^\dagger)^{N} | 0 \rangle ,
\label{H_2s}
\end{eqnarray}
where $0 \le \mu \le \nu$. Following the procedure in Eq.
(\ref{me_example}) it boils down to the expression
\begin{eqnarray}
t_{\alpha_1,\alpha_2,...,\alpha_{p-\mu};\beta_1,\beta_2,...,\beta_p}^{[\gamma_1,\gamma_2,...,\gamma_r]}
,  \label{H_t}
\end{eqnarray}
where in the scripts the number of $\alpha$'s, $\beta$'s, and
$\gamma$'s satisfy
\begin{eqnarray}
p + \frac{r}{2} \le \nu+1 \le p + r ,~~ p \ge \mu  \label{s_range} .
\end{eqnarray}
Below we compute the number of different $t$'s at given $\mu$, $p$,
and $r$.

In the case of the spherical shell model with rotational symmetry,
the single-particle space is generally written as $\{j_1, j_2, ...,
j_D\}$, each level $j_i$ has degeneracy $2 \Omega_i = 2 j_i + 1$.
The quantity $t$ (\ref{H_t}) is independent of the magnetic quantum
number $m$: $t_{...,{jm},...;...}^{[...]} =
t_{...,{jm'},...;...}^{[...]}$ for arbitrary $m$ and $m'$; similarly
in the $\beta$ and $\gamma$ indices. Hence in this case we could
write the quantity $t$ (\ref{H_t}) in ``occupation representation''
as
\begin{eqnarray}
t_{n^\alpha_1,n^\alpha_2,...,n^\alpha_D;n^\beta_1,n^\beta_2,...,n^\beta_D}^{[n^\gamma_1,n^\gamma_2,...,n^\gamma_D]}
,  \label{H_t_rot}
\end{eqnarray}
where $n^\alpha_i$ is the number of $j_i$'s (with arbitrary magnetic
quantum number $m_i$) present in the series
$\alpha_1,\alpha_2,...,\alpha_p$; similarly for $n^\beta_i$ and
$n^\gamma_i$. The following relations hold:
\begin{eqnarray}
\sum_i n^\alpha_i = p-\mu ,~~ \sum_i n^\beta_i = p ,~~ \sum_i
n^\gamma_i = r ,
\label{rel1} \\
0 \le n^\alpha_i + n^\beta_i + n^\gamma_i \le \Omega_i ,~~ 0 \le
n^\alpha_i , n^\beta_i , n^\gamma_i \le \Omega_i .     \label{rel2}
\end{eqnarray}
We count the number of solutions $(n^\alpha_i , n^\beta_i ,
n^\gamma_i)$ of Eqs. (\ref{rel1}) and (\ref{rel2}) to get the number
of different $t$'s at given $\mu$, $p$, and $r$. In practice, the
number of non-negative integer solutions satisfying Eq. (\ref{rel1})
is $C_{r+D-1}^{D-1} C_{p+D-1}^{D-1} C_{p-\mu+D-1}^{D-1}$, from which
we remove those violating Eq. (\ref{rel2}).

In practical calculations we usually truncate the many-body space up
to a maximum seniority $S = 2s$, thus the allowed values of the
quartet $(\nu, \mu, p, r)$ satisfy
\begin{eqnarray}
0 \le \nu \le s ,~~ 0 \le \mu \le \nu ,~~ \mu \le p \le \nu+1 ,
\nonumber \\
\nu + 1 - p \le r \le 2(\nu + 1 - p) .  \nonumber
\end{eqnarray}
For each allowed triplet $(\mu, p, r)$, we count the number of
different $t$'s (\ref{H_t}) based on Eqs. (\ref{rel1}) and
(\ref{rel2}). Then we sum the results for all possible triplets
$(\mu, p, r)$ to get the total number of $t$'s needed for a
calculation truncated at seniority $S=2s$. In Table
\ref{Table_num_t} we list the numbers for realistic nuclear
single-particle spaces. We see that it is indeed possible to store
the intermediate quantities $t$ (\ref{H_t}) within the memory of
modern computers, which have several gigabytes in a laptop and
several terabytes at a workstation.

In summary, the procedure of doing a realistic calculation for
even-even nuclei truncated at generalized seniority $S=2s$ is: 1.
Calculate the structure $v_\alpha$ (\ref{P_dag}) of the pair forming
the condensate (\ref{gs}), for protons and neutrons separately. 2.
Compute all the intermediate quantities $t$ (\ref{H_t}) based on the
recursive relations (\ref{t_rec}), and store the results in memory.
3. Construct the many-body space consisting of basis states with
fixed seniority up to $S$, additional truncation may be introduced
to further reduce the dimension. 4. Calculate the overlaps of the
basis and the matrices of operators (e.g. the Hamiltonian) in a way
similar to that of Eq. (\ref{me_example}). 5. Diagonalize the sparse
Hamiltonian matrix and calculate other observables.

A few comments are necessary. In step 1, the pair structure
$v_\alpha$ (\ref{P_dag}) could be determined by the conventional way
of minimizing energy, or by the recent method \cite{Jia_4} based on
the generalized density matrix that is much quicker in large model
spaces \cite{Jia_5}. The time cost of step 2 is not a problem at
all, because computing each $t$ (\ref{H_t}) needs only a few
multiplication operations according to Eq. (\ref{t_rec}), and the
process is fully parallelable.

Step 4 is very similar to that of the ``m-scheme'' shell model but
with two major differences. First, in the shell model a matrix
element vanishes unless the operator series $aa...a$ from the left
vector is the same (in one-to-one correspondence) as the series
$a^\dagger a^\dagger ... a^\dagger$ from the right vector, but here
they could differ in time-reversed pairs. Thus the Hamiltonian
matrix is less sparse compared with that of the shell model. Second,
the value of the shell model matrix element is simply $1$ (possibly
with a phase ``$-1$''), but here the matrix element boils down to a
specific $t$ that we should look up (for example by binary search)
in the memory. This process is also parallelable and the time cost
should not be a problem.

Even with the seniority truncation the many-body dimension becomes
very large in big single-particle spaces at high seniority. The
limiting factor in this situation should be the incapability of
diagonalizing the huge Hamiltonian matrix, as was the case with the
shell model. Thus in step 3 additional truncations beyond the
seniority truncation could be introduced to further reduce the
dimension of the many-body space. For example, in a two-major-shell
calculation we could make the popular restriction that there was a
maximum number $c$ of \emph{unpaired particles} that were allowed to
be excited to the upper major shell. Please note that the actual
number of particles on the upper shell could well exceed the number
$c$, because the pairing condensate also has components in the upper
shell (excitations due to pairing interaction). Another popular
truncation was to use the collective pairs with certain multiplicity
\cite{IBM_book, Chen_1997, Zhao_2000, Yoshinaga_2000} if the
corresponding multipole-multipole interaction in the Hamiltonian was
believed to be significant. Usually for the low-lying states the $D$
pair (quadrupole pair with angular momentum two) is the most
important one.

\section{Summary     \label{Sec_sum}}

In conclusion, we propose a scheme for doing practical calculations
with generalized seniority. The method utilizes the huge memory
capabilities of modern computers by calculating and storing a set of
intermediate quantities to reduce the total computing time costs.
The requirements (memory and speed) and performance of the algorithm
are analyzed in detail.

The limiting factor of the method is still the dimension of the
many-body space. Even with the seniority truncation, in large
single-particle spaces the many-body space may still become
intractably huge at relatively high seniority. Thus additional
truncations or restrictions on the unpaired particles may be
necessary. Convergence should be reached for a specific observable
with respect to the cutoffs of the truncations should the seniority
results accurately reproduce the shell model results.

The current scheme of seniority calculations is similar in
programming to that of the shell model. Thus it should be relatively
easy to modify the existing well-developed ``m-scheme'' shell model
codes to get a good
seniority code. Mature techniques used there could be adopted.\\

Support is acknowledged from the startup funding for new faculty
member in University of Shanghai for Science and Technology. Part of
the calculations is done at the High Performance Computing Center of
Michigan State University.

\begin{table}
\caption{\label{Table_num_t}  Number of different $t$'s (\ref{H_t})
needed for calculations truncated at generalized seniority $S=2s$ in
realistic single-particle spaces. The row labels represent
single-particle spaces taken between two magic numbers; for example,
``$8 \sim 50$'' represents the space $\{ 0d_{\frac{5}{2}},
0d_{\frac{3}{2}}, 1s_{\frac{1}{2}}, 0f_{\frac{7}{2}},
0f_{\frac{5}{2}}, 1p_{\frac{3}{2}}, 1p_{\frac{1}{2}},
0g_{\frac{9}{2}} \}$. For the suffix of numbers we have $1G = \chi M
= \chi^2 k = \chi^3$, with $\chi = 1024 = 2^{10}$, following the
convention in computers. }
\begin{tabular}{|c|c|c|c|c|c|c|}
  \hline
  ~            & $s = 1$ & $s = 2$ & $s = 3$ & $s = 4$ & $s = 5$ & $s = 6$ \\
  \hline
  $20 \sim 50$ & 846 & 7.42k & 39.2k & 135k & 317k & 537k \\
  $50 \sim 82$ & 846 & 7.42k & 39.5k & 140k & 351k & 652k \\
  $82 \sim 126$ & 1.63k & 20.8k & 161k & 848k & 3.18M &  9.29M \\
  \hline
  $8 \sim 50$ & 4.28k & 79.9k & 845k & 5.54M & 25.2M & 81.4M \\
  $20 \sim 82$ & 10.2k & 302k & 5.10M & 58.1M & 468M & 2.73G  \nonumber \\
  $50 \sim 126$ & 14.7k & 527k & 10.9M & 152M & 1.48G & 11.2G \\
  \hline
  $8 \sim 82$ & 26.9k & 1.23M & 32.9M & 571M & 6.76G & 60.2G \\
  $28 \sim 126$ & 47.1k & 2.88M & 104M & 2.42G & 40.8G &  512G \\
  \hline
  $0 \sim 126$ & 201k & 24.6M & 1.70G & 77.7G & --- & --- \\
  \hline
\end{tabular}
\end{table}

\end{document}